\documentclass[10pt]{iopart}
\usepackage{graphicx}
\usepackage{iopams}
\begin{document}
\newcommand{\ket}[1]{| #1 \rangle} % for Dirac bras
\newcommand{\bra}[1]{\langle #1 |} % for Dirac kets
\newcommand{\braket}[2]{\langle #1 \vphantom{#2} | #2 \vphantom{#1} \rangle} % for Dirac brackets

\title{Local and nonlocal probabilistic cloning of qubit states via partially entangled twin
photons}
\author{N Cisternas $^1$, G Araneda$^2$,
O. Jim\'{e}nez$^3$ and  A Delgado$^{4,5,6}$}
\address{$^1$ Van der Waals-Zeeman Institute for Experimental Physics, Universiteit van Amsterdam, Science Park 904,1098 XH Amsterdam, The Netherlands}
\address{$^2$ Institut f\"{u}r Experimentalphysik,Universit\"{a}t Innsbruck, Technikerstrasse 25, 6020 Innsbruck, Austria}
\address{$^3$Departamento de F\'isica, Facultad de Ciencias B\'asicas, Universidad de Antofagasta, Casilla 170, Antofagasta, Chile}
\address{$^4$Center for Optics and Photonics, Universidad de Concepci\'{o}n, Casilla 4012, Concepci\'{o}n, Chile}
\address{$^5$MSI-Nucleus on Advanced Optics, Universidad de Concepci\'on, Casilla 160-C, Concepci\'on, Chile}
\address{$^6$Departamento de F\'isica, Universidad de Concepci\'on, Casilla160-C, Concepci\'on, Chile}
\ead{gabriel.araneda-machuca@uibk.ac.at}
\begin{abstract}
Recently, the nonlocal optimal probabilistic cloning (NLOPC) of two non-orthogonal qubit states has been proposed [Phys. Rev. A {\bf 86}, 052332 (2012)] by means of an experimental setup based on a pair of twin photons in a maximally entangled state.  Here we study the performance of the NLOPC protocol when implemented via a partially entangled state. We show that the errors introduced by the use of partial entanglement can be undone by applying a quantum state discrimination process. Since quantum state discrimination is a probabilistic process the correction succeeds with a certain probability but produces perfect clones. We also studied how modified the setup to produce local copies.

\end{abstract}

%\submitto{}
\maketitle
\ioptwocol
\section{Introduction}

The laws of quantum mechanics forbid the perfect cloning of unknown quantum states \cite{Wootters,Dieks,Milonni,Mandel}. This important result, known as the no-cloning theorem, bases on the linearity of the quantum-mechanical transformations and has been subject of continuous research in the last decades. In particular, the problem of approximate cloning has been exhaustively explored \cite{Buzek,Werner,Keyl,Alber}. This is mainly due to the fact that it imposes a limit on our capability to manipulate and broadcast quantum information and because it can be used as an efficient eavesdropping attack on quantum key distribution (QKD) \cite{Scarani}.

The impossibility of  generating perfect copies
imposed by the no-cloning theorem can be relaxed by resorting a non deterministic process. This kind of process is called probabilistic cloning, where a set of known linearly independent nonorthogonal quantum states can be cloned perfectly at the expense of succeeding with probability less than one \cite{Duan,Duan2}. This kind of cloning is of particular interest when testing the security of the so called  B92 QKD protocol \cite{B92}, which involves only two non-orthogonal linearly independent states. Other applications and implementations are studied in \cite{Rev} and \cite{appli}

 Recently, this probabilistic cloning scheme  has been successfully implemented via nuclear magnetic resonance \cite{Hongwei} in the case of generating two clones of a single qubit state. This demonstration, however, does not allow the broadcasting of the cloned states. This led us to propose the implementation of an optimal  probabilistic cloning based on entangled twin photons \cite{Araneda}. This proposal is based on linear optics together with a pair of maximally entangled twin photons generated via spontaneous parametric down-conversion. The states to be cloned are encoded in the polarization degree of freedom of one of the twin photons. After a sequence of measurements the state to be cloned as well as its perfect clone are encoded in the polarization and spatial degree of freedom of the resting photon respectively. The former photon is destroyed in a photo-detection process. Thereby, the clones are produced in a non-local way and can be broadcasted.

Sources of polarized entangled pairs of photons have been shown to have very good fidelities when comparing with Bell states \cite{photons}, but additional loses can appear when distributing the entangled pairs at large distances, and also other fluctuations can make the state imperfect. Therefore even if the source is close to perfect the state shared by  the different parties  is not necessary close to perfect. This is a serious issue, since perfect clones are created only when the parties share a perfect entangled state. In this work we present an example of how is still possible to have high fidelity copies even if the entanglement is not high quality. We study the performance of the NLOPC process when cloning a state from a known set of linearly independent nonorthogonal states, implemented via a partially entangled state of a pair of twin photons. We show that the errors introduced by the use of partial entanglement can be undone by applying a quantum state discrimination process to the photon which encodes original and copy states. This process is applied after the NLOPC protocol has been carried out. The correction procedure is based on the application of unambiguous state discrimination and  does not requires information about the state undergoing the cloning process \cite{DelgadoC,DelgadoD}. This procedure is complete determined by the Schmidt coefficients of the partially entangled state and the measurement results of the cloning process. Since quantum state discrimination is a probabilistic process the correction succeeds with a certain probability. The clones emerging from the correction procedure are perfect but the overall success probability of generating perfect clones is thus reduced by a factor given by the optimal probability of unambiguously discriminating two nonorthogonal states, that is the Ivanovic-Dieks-Peres limit. The states to be discriminated are defined by the Schmidt coefficients of the partially entangled quantum channel. 

We also study the possibility of local probabilistic cloning (LPC) where the states to be cloned are encoded in the polarization degree of freedom of one of the twin photons while a perfect clone is generated and encoded in the spatial degree of freedom of the same photon. Thus, the clones are produced locally. The degrees of freedom of the resting photon are used as ancillary systems to implement conditional transformations. The experimental setup for LPC turns out to be simpler than in the case of NLOPC but the success probability of the LPC is the half of the success probability of NLOPC. We also study the case local probabilistic cloning via partially entangled twin photons. 

This article is organized as follows: in Section \ref{SECTIONII} we summarize briefly the the optimal probabilistic cloning process. In Section \ref{SECTIONIIIb} we present the experimental setup for implementing the LPC. Section \ref{SECTIONIV} is devoted to study the performance of the proposed setup for LPC when the entanglement it is not maximal. In Section \ref{SECTION5} we analyze the NLOPC via a partially entangled state and show that the addition of an unambiguous state discrimination stage corrects the errors generated by the partial entanglement at the expense of reducing the overall success probability of generating perfect clones. In Section \ref{SECTIONV} we summarize our results.

%%%%%%%%%%%%%%%%%%%%%%%%%%%%%%%%%%%%%%%%%%%%%%%%%%%%%%

\section{Optimal probabilistic quantum cloning machine}
\label{SECTIONII}

The probabilistic quantum cloning machine is constructed by concatenating a unitary transformation $U$ and a projective measurement \cite{Duan,Duan2}. The action of the unitary transformation is
\begin{eqnarray}
U|\psi_\pm\rangle_o|\Sigma\rangle_{ca}=\sqrt{p}|\psi_\pm\rangle_o|\psi_\pm\rangle_c|0\rangle_a
+\sqrt{1-p}|\Phi\rangle_{oc}|1\rangle_a,\nonumber\\
&
\label{U}
\end{eqnarray}
where $|\psi_\pm\rangle_o$ are two nonorthogonal states encoded in the original system $o$. These are the states to be cloned. The perfect clone of each one of these states is encoded in the copy system $c$ as $|\psi_\pm\rangle_c$. Additionally, the transformation requires a third ancillary system $a$ spanned by the orthogonal states $|0\rangle_a$ and $|1\rangle_a$. Initially, systems $c$ and $a$ are described by a joint blank state $|\Sigma\rangle_{ca}$. After the unitary transformation $U$ has been carried out a measurement on system $a$ projects it onto state $|0\rangle_a$. Systems $o$ and $c$ are then described by a product state formed by the original state and its clone. This succeeds with an optimal probability $p_{opt}$ given by
\begin{eqnarray}
p_{opt}=\frac{1}{1+|\langle\psi_+|\psi_-\rangle|}.
\label{success-probability}
\end{eqnarray}
In case of projecting system $a$ onto state $|1\rangle_a$ systems $o$ and $c$ are left in an entangled state $|\Phi\rangle_{oc}$, which does not correspond to a cloning process. Considering the states to be cloned as $|\psi_\pm\rangle_o=\cos(\theta/2)|0\rangle_o\pm\sin(\theta/2)|1\rangle_o$,  the action of the unitary transformation $U$ onto the basis states $|0\rangle_o$ and $|1\rangle_o$ is given by
\begin{eqnarray}
U|0\rangle_{o}|\Sigma \rangle_{ca}& =  (\tilde{\alpha}  |0\rangle_{a} + \tilde{\beta } |1\rangle_{a})(\alpha |0\rangle_{o}|0\rangle_{c} + \beta |1\rangle_{o}|1\rangle_{c}), \\
U|1\rangle_{o}|\Sigma \rangle_{ca}& =  |0\rangle_{a} \frac{1}{\sqrt{2}}  ( |0\rangle_{o}|1\rangle_{c} +  |1\rangle_{o}|0\rangle_{c}),
\end{eqnarray}
where coefficients $\alpha, \beta, \tilde\alpha$ and $\tilde\beta$ are
\begin{eqnarray}
\alpha&=&\frac{1}{\sqrt{1+\tan^4(\theta/2)}},~~\beta=\frac{\tan^2(\theta/2)}{\sqrt{1+\tan^4(\theta/2)}},
\label{alpha-beta}\\
\tilde\alpha&=&\sqrt{\frac{1+\tan^4(\theta/2)}{2}},~~\tilde\beta=\sqrt{\frac{1-\tan^4(\theta/2)}{2}}.
\label{alpha-beta-tilde}
\label{SETTINGS}
\end{eqnarray}
In this case the success probability, according to Eq. (\ref{success-probability}), is given by
\begin{eqnarray}
p_{opt}=\frac{1}{1+|\cos(\theta)|},
\label{success-probability-particular}
\end{eqnarray}
and the state $|\Phi\rangle_{oc}$ becomes
\begin{eqnarray}
|\Phi\rangle_{oc}=\alpha|0\rangle_o|0\rangle_c+\beta|1\rangle_o|1\rangle_c.
\end{eqnarray}
This does not depend on the particular state being cloned. 

A simple figure of  merit to compare the performance of different cloning schemes is the multiplication of the fidelity of the cloned states respect to the input states times the probability to get it, $f\times p$. In the scheme that we study the fidelity is unitary and the probability is given by eq. \ref{success-probability-particular}, then $f\times p=1/(1+\cos\theta)$. We can compare this with the performance of the universal cloning machine, see for example \cite{Scarani}, which is a deterministic process, but with limited fidelity equal to $5/6$, in which case $f\times p=5/6$. It is straight forward to see that our scheme of probabilistic cloning overperforms universal cloning when $|\cos\theta|< 1/5$ considering this particular figure of merit.

Analytical solutions for the problem of cloning larger sets of linearly independent states are known in cases where the states to be cloned have a high degree of symmetry \cite{DelgadoA,DelgadoB}.

\section{Local probabilistic cloning via a maximally entangled polarization state}
\label{SECTIONIIIb}

%%%%%%%%%%%%%%%%%%%%%%%%%%%%%%%%%%%%%%%%%%%%%%%%%%%%%%%%%%%

\begin{figure*}[ht]
\centerline{
\includegraphics[width=0.9\textwidth]{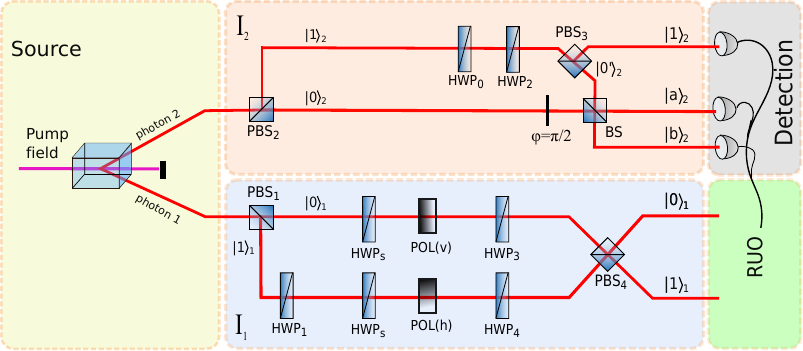}}
\caption{Experimental setup for implementing local probabilistic cloning. Two photons are generated in a maximally entangled polarization state followed by two polarizer beam splitters, $PBS_{1}$ and $PBS_{2}$, which generate two propagation paths for each photon. Then, half-wave plates $HWP_{0}$ and $HWP_{1}$ prepare the initial state needed for the process. The state to be cloned is encoded in the polarization of photon 1 by means of half-wave plates $HWP_{s}$. The process of probabilistic cloning is carried out by interferometers $I_{1}$ and $I_{2}$. After photo-detection of photon 2 at output ports of interferometer $I_2$ a clone is generated in the spatial degree of freedom of photon 1.  Conditional on the measurement outcome a Reconstruction Unitary Operation (RUO) is applied on photon 1. This latter transformation does not require information about the particular state undergoing the cloning process.}
\label{Setup}
\end{figure*}

The experimental setup proposed for implementing the local probabilistic cloning of single-qubit states is illustrated in Fig. (\ref{Setup}). This requires the use of a source for generating a maximally entangled polarization state of two photons. This state can be  generated via spontaneous parametric down-conversion of type-I using the scheme presented in \cite{Kwiat} and is of the form
\begin{eqnarray}
|\Psi_s\rangle_{12}=\frac{1}{\sqrt{2}}(|h\rangle_1|h\rangle_2+|v\rangle_1|v\rangle_2),
\label{optimal-resource}
\end{eqnarray}
where $|h\rangle_i$ and $|v\rangle_i$ describe horizontal and vertical polarization for photon $i=1,2$. 

Photon $1$ is directed toward interferometer $I_1$ which is defined by polarizing beam splitters $PBS_1$ and $PBS_4$. States $|0\rangle_1$ and $|1\rangle_1$ describe the spatial state of photon $1$ propagating at the upper or lower arm of the interferometer, correspondingly. Photon $2$ enters into interferometer $I_2$ whose two input ports are defined by polarizing beamsplitter $PBS_2$. The three output ports of $I_2$ are generated by polarizing beam splitter $PBS_3$ and beam splitter $BS$. States $|1\rangle_2$ and $|0\rangle_2$ describe the spatial state of the photon $2$ propagating at the upper or lower arm of interferometer $I_2$, respectively. This interferometer is lossy since polarizing beamsplitter $PBS_3$ allows vertically polarized photons to leave the interferometer. 

The state of the twin photons after polarizing beam splitters $PBS_1$ and $PBS_2$ is given by
\begin{eqnarray}
|\Psi_0\rangle_{12}=\frac{1}{\sqrt{2}}(|h\rangle_1|0\rangle_1|h\rangle_2|0\rangle_2+|v\rangle_1|1\rangle_1|v\rangle_2|1\rangle_2).
\end{eqnarray}
On the lower arm of interferometer $I_1$ a half-wave plate $HWP_1$ is located. This transforms vertically polarized photons into horizontally polarized photons. Thereby, state $|\Psi_0\rangle_{12}$ becomes
\begin{eqnarray}
|\Psi_1\rangle_{12}=\frac{1}{\sqrt{2}}|h\rangle_1(|0\rangle_1|h\rangle_2|0\rangle_2+|1\rangle_1|v\rangle_2|1\rangle_2),
\end{eqnarray}
where the polarization of photon $1$ is disentangled from the resting degrees of freedom.

Both arms of interferometer $I_1$ contain a half-wave plate $HWP_s$ which allows to encode in the polarization of photon $1$ the states to be cloned. The tilting angle of both half-wave plates $HWP_s$ is equal and depends on the state to be encoded. After these half-wave plates the state of photon $1$ is given by
\begin{eqnarray}
|\Psi_2\rangle_{12}=\frac{1}{\sqrt{2}}|\psi_\pm\rangle_1(|0\rangle_1|h\rangle_2|0\rangle_2+|1\rangle_1|v\rangle_2|1\rangle_2),
\end{eqnarray}
where $|\psi_\pm\rangle_1=a|h\rangle_1\pm b|v\rangle_1$ with $a=\cos(\theta/2)$ and $b=\sin(\theta/2)$. Thereby, the polarization degree of freedom of photon $1$ plays the role of the system $o$ while the spatial degree of freedom plays the role of the system $c$, which will encodes the clones, and photon $2$ plays the role of the ancilla system $a$.

After half-wave plates $HWP_s$ polarizers $POL(v)$ and $POL(h)$ are located at upper and lower arm respectively. These filter the vertical and horizontal polarization of photon $1$ conditional on the propagation path. After these polarizers the normalized state of photons $1$ and $2$ is given by
\begin{eqnarray}
|\Psi_3\rangle_{12}=a|h\rangle_1|1\rangle_1|v\rangle_2|1\rangle_2\pm b|v\rangle_1|0\rangle_1|h\rangle_2|0\rangle_2.
\label{Replacement1}
\end{eqnarray}
This state is generated with a probability of $1/2$. This state is modified with half-wave plates $HWP_3$ and $HWP_4$. The former transforms vertically polarized photons propagating at the upper arm of $I_1$ into an equally weighted superposition of vertical and horizontal polarization. The latter maps the horizontal polarization state of photon $1$ propagating at the lower arm of $I_1$ onto the polarization superposition $\alpha|h\rangle_1+\beta|v\rangle_1$, with $\alpha$ and $\beta$ given by Eq. (\ref{alpha-beta}). The new state is
\begin{eqnarray}
|\Psi_4\rangle_{12}&=&a(\alpha|h\rangle_1+\beta|v\rangle_1)|1\rangle_1|v\rangle_2|1\rangle_2
\nonumber\\
&\pm& b\frac{1}{\sqrt{2}}(|h\rangle_1+|v\rangle_1)|0\rangle_1|h\rangle_2|0\rangle_2.
\end{eqnarray}
Finally, photons propagating in interferometer $I_1$ exit through polarizing beam splitter $PBS_4$. After this the joint state of photons $1$ and $2$ is given by
\begin{eqnarray}
|\Psi_5\rangle_{12}&=&a(\alpha|h\rangle_1|0\rangle_1+\beta|v\rangle_1|1\rangle_1)|v\rangle_2|1\rangle_2
\nonumber\\
&\pm& b\frac{1}{\sqrt{2}}(|h\rangle_1|1\rangle_1+|v\rangle_1|0\rangle_1)|h\rangle_2|0\rangle_2.
\end{eqnarray}
The state of photon $2$ is modified by interferometer $I_2$. Here, half-wave plate $HWP_{0}$ changes the polarization of photons propagating at the upper arm from vertical to horizontal, that is
\begin{eqnarray}
|\Psi_6\rangle_{12}&=a(\alpha|h\rangle_1|0\rangle_1+\beta|v\rangle_1|1\rangle_1)|h\rangle_2|1\rangle_2
\nonumber\\
&\pm b\frac{1}{\sqrt{2}}(|h\rangle_1|1\rangle_1+|v\rangle_1|0\rangle_1)|h\rangle_2|0\rangle_2.
\end{eqnarray}

Half-wave plate $HWP_2$ transforms the horizontal polarization state of photons propagating on the upper arm into a polarization superposition of the form $\tilde\alpha|h\rangle_2+\tilde\beta|v\rangle_2$, with $\tilde\alpha$ and $\tilde\beta$ given by Eq. (\ref{alpha-beta-tilde}). Photons propagating at the lower arm of $I_2$ acquire an additional phase of $\pi/2$. We obtain
\begin{eqnarray}
|\Psi_7\rangle_{12}&=&a(\alpha|h\rangle_1|0\rangle_1+\beta|v\rangle_1|1\rangle_1)(\tilde\alpha|h\rangle_2+\tilde\beta|v\rangle_2)|1\rangle_2
\nonumber\\
&\pm& ib\frac{1}{\sqrt{2}}(|h\rangle_1|1\rangle_1+|v\rangle_1|0\rangle_1)|h\rangle_2|0\rangle_2.
\end{eqnarray}
Thereafter, photons propagating at the upper arm are directed toward polarizing beam splitter
$PBS_3$, which generates the state
\begin{eqnarray}
|\Psi_8\rangle_{12}&=&a(\alpha|h\rangle_1|0\rangle_1+\beta|v\rangle_1|1\rangle_1)(\tilde\alpha|h\rangle_2|0'\rangle_2+\tilde\beta|v\rangle_2|1\rangle_2)
\nonumber\\
&\pm& ib\frac{1}{\sqrt{2}}(|h\rangle_1|1\rangle_1+|v\rangle_1|0\rangle_1)|h\rangle_2|0\rangle_2.
\end{eqnarray}
Finally, photons emerging from polarizing beam splitter $PBS_3$ with horizontal polarization are merged with photons propagating at the lower arm of $I_2$ at beam splitter $BS$. This transforms states $|0\rangle_2$ and $|0'\rangle_2$ into states $(1/\sqrt{2})(|a\rangle_2+i|b\rangle_2)$ and $(1/\sqrt{2})(i|a\rangle_2+|b\rangle_2)$, respectively, and leads to the joint state

\begin{eqnarray}
|\Psi_9\rangle_{12} & = & a\tilde{\beta}|\phi^{(1)}\rangle_1|v\rangle_2|1\rangle_2+\frac{i}{\sqrt{2}}|\phi^{(2)}\rangle_1|h\rangle_2|a\rangle_2
\nonumber\\
&+&\frac{1}{\sqrt{2}}|\phi^{(3)}\rangle_1|h\rangle_2|b\rangle_2,
\label{CLONING1}
\end{eqnarray}
where the non normalized states of photon $1$ entering in the previous expression are given by
\begin{eqnarray}
|\phi^{(1)}\rangle_1&=&\alpha |h\rangle_{1}|0\rangle_{1}+\beta |v\rangle_{1}|1\rangle_{1},
\nonumber\\
|\phi^{(2)}\rangle_1&=&a\tilde{\alpha}(\alpha |h\rangle_{1}|0\rangle_{1}+\beta |v\rangle_{1}|1\rangle_{1}) \nonumber \\
&\pm & b\frac{1}{\sqrt{2}}(|h\rangle_{1}|1\rangle_{1}+ |v\rangle_{1}|0\rangle_{1}),
\nonumber\\
|\phi^{(3)}\rangle_1&=&a\tilde{\alpha}(\alpha |h\rangle_{1}|0\rangle_{1}+\beta |v\rangle_{1}|1\rangle_{1}) \nonumber \\
&\mp & b\frac{1}{\sqrt{2}}(|h\rangle_{1}|1\rangle_{1}+ |v\rangle_{1}|0\rangle_{1}).
\label{CLONING2}
\end{eqnarray}

According to state $|\Psi_9\rangle_{12}$ a photo-detection at the output ports of interferometer $I_2$ projects photon $1$ onto three possible states. A photo-detection of photon $2$ in path $1$ projects photon $1$ onto state $|\phi^{(1)}\rangle_1$. This state is obtained independently of the state $|\psi_\pm\rangle_1$ to be cloned and represents the failure of the cloning attempt. A photo-detection on path $a$ projects photon $1$ onto the state $|\phi^{(2)}\rangle_1$. Considering Eqs. (\ref{alpha-beta}) and (\ref{alpha-beta-tilde}) this latter state can be cast in the form
\begin{eqnarray}
\frac{1}{\sqrt{2}a}(a|h\rangle_1\pm b|v\rangle_1)(a|0\rangle_1\pm b|1\rangle_1),
\label{cloning-correct}
\end{eqnarray}
which describes a successful cloning process of the states $|\psi_\pm\rangle_1$.  Finally, a photo-detection on path $b$ projects photon $1$ onto the state $|\phi^{(3)}\rangle_1$,  which can be cast in the form
\begin{eqnarray}
\frac{1}{\sqrt{2}a}(a|h\rangle_1\mp b|v\rangle_1)(a|0\rangle_1\mp b|1\rangle_1).
\label{cloning-wrong}
\end{eqnarray}
Here the attempt of cloning states $|\psi_\pm\rangle_1$ leads to the clones of states $|\psi_\mp\rangle_1$. However, this error can be corrected without the knowledge of the particular state being cloned and conditional to a photo-detection on path $b$. The correction stage is implemented by applying a recovery unitary operator (RUO). This is implemented by placing a retarder plate on path $1$ of photon $1$ followed by a half-wave plate on each path available for this photon. Both half-wave plates interchange horizontal and vertical polarization. After this correction stage the state (\ref{cloning-wrong}) is transformed into the state (\ref{cloning-correct}). 

Considering photo-detections in paths $a$ and $b$ of photon $2$ and the conditional correction stage on photon $1$, the success probability $p_s$ of the cloning process here proposed is given by 
\begin{eqnarray}
p_s=1-|a\tilde\beta|^2
\end{eqnarray}
or equivalently
\begin{eqnarray}
p_s=p_{opt},
\end{eqnarray}
with $p_{opt}$ given by Eq. (\ref{success-probability-particular}). However, the filtering process, which is carried out after encoding the state to be cloned, reduces the success probability by a factor $1/2$. Thereby, the implementation of probabilistic cloning via a maximally entangled state here proposed achieves the total success probability $p_{total}^{me}$ given by
\begin{eqnarray}
p_{total}^{me}=p_{opt}/2.
\end{eqnarray}
Thereby, our cloning process turns out to be suboptimal. 

%The success probability of our scheme can be increased to the value of $p_{opt}$. This can be done by replacing the polarizers in the interferometer $I_1$ with polarizing beam splitters. Photons at the upper arm with vertical polarization and at the lower arm with horizontal polarization continue to propagate in $I_1$. Photons at the upper arm with horizontal polarization and at the lower arm with vertical polarization are directed toward an interferometer $I_{1'}$. Photons with horizontal (vertical) polarization propagates at the upper (lower) arm of this new interferometer. At each arm of $I_{1'}$ the polarization of the photons is inverted and thereafter they are manipulated with the same transformations entering in the interferometer $I_1$. This allows us to use all pairs of photons produced by the source increasing the success probability of the cloning process from $p_{opt}/2$ to $p_{opt}$. However, this solution to achieve the optimal cloning probability has a drawback. A photo-detection at the exit ports of interferometer $I_2$ 
%leads to a superposition of two mutually orthogonal states of photon $1$. Each one of these describes a single photon emerging from interferometer $I_1$ or $I_{1'}$, being the probability of each of these events equal to $1/2$. Nevertheless, this photo-detection indicates if the cloning attempt fails, succeeds or requires the application of a RUO on both interferometers $I_1$ and $I_{1'}$ simultaneously.

\section{Local probabilistic cloning via partially entangled polarization states}
\label{SECTIONIV}

\begin{figure}
\includegraphics[width=0.45\textwidth]{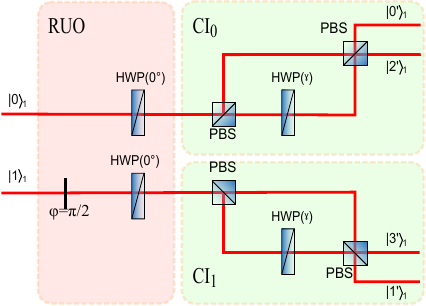}
\caption{Experimental setup for implementing the reconstruction unitary operation RUO and conditional interferometers $CI_0$ and $CI_1$. RUO is implemented with a retarder plate followed by a half-wave plate (HWP) on each path available for photon 1. These wave plates are set to exchange horizontal and vertical polarization. Input and output ports of interferometers $CI_0$ and $CI_1$ are defined by polarizing beam splitters (PBS). Each interferometer has in one arm a half-wave plate, which allows for polarization transformations (\ref{U0}) and (\ref{U1}) conditional to the propagation path of photon 1. The tilting angle $\gamma$ of the half-wave plates depends on the Schmidt coefficients of the quantum channel through Eq. (\ref{Gamma}).}
\label{Transformation}
\end{figure}

In the previous section we have proposed an experimental setup for implementing the probabilistic cloning of two nonorthogonal single-qubit states. This was based on the use of the maximally entangled two-photon polarization state $|\Psi_s\rangle_{12}$ of Eq. (\ref{optimal-resource}). Here we show that it is possible to use a known partially entangled polarization state to implement the probabilistic cloning. This, however, leads to a reduction of the success probability.

Let us consider a partially entangled polarization state of photons $1$ and $2$, such as
\begin{eqnarray}
|\tilde\Psi_s\rangle_{12}=c_h|h\rangle_1|h\rangle_2+c_v|v\rangle_1|v\rangle_2.
\label{partial-resource}
\end{eqnarray}
For simplicity coefficients $c_h$ and $c_v$ are chosen as real numbers and such that $c_h\le c_v$. This is the input state for the experimental setup depicted in Fig. \ref{Setup}. After the filtering stage the twin photons are described by the state
\begin{eqnarray}
|\tilde\Psi_3\rangle_{12}&=\frac{c_va}{\sqrt{N}}|h\rangle_1|1\rangle_1|v\rangle_2|1\rangle_2\pm \frac{c_hb}{\sqrt{N}}|v\rangle_1|0\rangle_1|h\rangle_2|0\rangle_2,\nonumber\\
&
\end{eqnarray}
where $N=(c_va)^2+(c_hb)^2$ is the probability of photon $1$ to pass the filtering process. After the action of interferometers $I_1$ and $I_2$ the joint state of photons $1$ and $2$ is given by
\begin{eqnarray}
|\tilde\Psi_9\rangle & = & \frac{ac_v\tilde{\beta}}{\sqrt{N}}|\tilde\phi^{(1)}\rangle_1|v\rangle_2|1\rangle_2+\frac{i}{\sqrt{2N}}|\tilde\phi^{(2)}\rangle_1|h\rangle_2|a\rangle_2
\nonumber\\
&+&\frac{1}{\sqrt{2N}}|\tilde\phi^{(3)}\rangle_1|h\rangle_2|b\rangle_2,
\end{eqnarray}
where the non normalized states of photon $1$ entering in the previous expression are given by
\begin{eqnarray}
|\tilde\phi^{(1)}\rangle_1&=&\alpha |h\rangle_{1}|0\rangle_{1}+\beta |v\rangle_{1}|1\rangle_{1},
\nonumber\\
|\tilde\phi^{(2)}\rangle_1&=&ac_v\tilde{\alpha}(\alpha |h\rangle_{1}|0\rangle_{1}+\beta |v\rangle_{1}|1\rangle_{1}) \nonumber \\
&\pm & bc_h\frac{1}{\sqrt{2}}(|h\rangle_{1}|1\rangle_{1}+ |v\rangle_{1}|0\rangle_{1}),
\nonumber\\
|\tilde\phi^{(3)}\rangle_1&=&ac_v\tilde{\alpha}(\alpha |h\rangle_{1}|0\rangle_{1}+\beta |v\rangle_{1}|1\rangle_{1}) \nonumber \\
&\mp & bc_h\frac{1}{\sqrt{2}}(|h\rangle_{1}|1\rangle_{1}+ |v\rangle_{1}|0\rangle_{1}).
\end{eqnarray}
These expressions are obtained by replacing $a$ and $b$ in Eqs. (\ref{CLONING1}) and (\ref{CLONING2}) with $ac_v/\sqrt{N}$ and $bc_h/\sqrt{N}$, respectively. State $|\tilde\phi^{(1)}\rangle_1$ describes the failure of the cloning process. Considering Eqs. (\ref{alpha-beta}) and (\ref{alpha-beta-tilde}), the state $|\tilde\phi^{(2)}\rangle_1$ can be cast in the form
\begin{eqnarray}
|\tilde\phi^{(2)}\rangle_1&=&\frac{1}{\sqrt{2}a}[a(c_va|h\rangle_1\pm c_hb|v\rangle_1)|0\rangle_1
\nonumber\\
&\pm& b(c_ha|h\rangle_1\pm c_vb|v\rangle_1)|1\rangle_1].
\end{eqnarray}
According to this expression coefficients $c_h$ and $c_v$, which characterize the partially entangled state used as resource, lead to an error in the cloning process. This error can be, however, corrected. 

Let us consider the following two pairs of nonorthogonal polarization states
\begin{eqnarray}
|\alpha_{\pm}\rangle_1|0\rangle_1&=&c_v|h\rangle_1|0\rangle_1\pm c_h|v\rangle_1|0\rangle_1,
\label{ALFAS}\\
|\tilde\alpha_{\pm}\rangle_1|1\rangle_1&=&c_h|h\rangle_1|1\rangle_1\pm c_v|v\rangle_1|1\rangle_1.
\label{ALFAST}
\end{eqnarray}
These allow to cast the state $|\tilde\phi^{(2)}\rangle_1$ in the form
%\begin{eqnarray}
%|\tilde\phi^{(2)}\rangle_1=\frac{1}{\sqrt{2}a}\left \{ a\left [ \left (\frac{a}{2}\pm\frac{b}{2}\right)\ket{\alpha_+}_1+\left(\frac{a}{2}\mp\frac{b}{2}\right)\ket{\alpha_-}_1\right] \ket{0}_1
%\right \}
%\nonumber\\
%\frac{1}{\sqrt{2}a}\left \{ \pm b\left [ \left(\frac{a}{2}\pm\frac{b}{2}\right)\ket{\tilde\alpha_+}_1+\left(\frac{a}{2}\mp\frac{b}{2}\right)\ket{\tilde\alpha_-}_1\right] \ket{1}_1
%\right \}
%\end{eqnarray}
%with $d_{\pm}=(a\pm b)/2$.
\begin{eqnarray}
|\tilde\phi^{(2)}\rangle_1 &=\frac{1}{\sqrt{2}a}\left \{ a\left [ d_\pm\ket{\alpha_+}_1+d_\mp\ket{\alpha_-}_1\right] \ket{0}_1
\right.
\nonumber\\
& \left.\pm b\left [ d_\pm\ket{\tilde\alpha_+}_1+d_\mp\ket{\tilde\alpha_-}_1\right] \ket{1}_1
\right \},
\label{Wrong-state}
\end{eqnarray}
with $d_{\pm}=(a\pm b)/2$. If we could implement the following transformations
\begin{eqnarray}
|\alpha_{\pm}\rangle_1|0\rangle_1&\rightarrow&\frac{1}{\sqrt{2}}(|h\rangle_1|0\rangle_1\pm |v\rangle_1|0\rangle_1),
\label{TRANSFORMATION1}\\
|\tilde\alpha_{\pm}\rangle_1|1\rangle_1&\rightarrow&\frac{1}{\sqrt{2}}(|h\rangle_1|1\rangle_1\pm |v\rangle_1|1\rangle_1),
\label{TRANSFORMATION2}
\end{eqnarray}
then these would map state $|\tilde\phi^{(2)}\rangle_1$ onto state of Eq. (\ref{cloning-correct}), which describes a perfect cloning process. Since these transformations map nonorthogonal states onto orthogonal states they cannot be carried out by a unitary transformation. However, they can be carried out probabilistically. This class of transformations has been extensively studied in the context of quantum state discrimination \cite{Ivanovic87,Chefles98-2,Chefles98-1}. Here the problem consists in the discrimination or identification of states belonging to a set of nonorthogonal states. Many discrimination strategies have been proposed depending on the figure of merit to be optimized \cite{Chefles98-1,Croke06}. In particular, unambiguous state discrimination \cite{Chefles98-1,DelgadoE} allows to identify linearly independent nonorthogonal states by mapping them onto orthogonal states, which can be thereafter distinguished with certainty. The main cost of this strategy is the introduction of an inconclusive outcome characterized by a certain probability of ocurrence, which can be minimized. In the case of discriminating between two nonorthogonal states $\ket{\psi_1}$ and $\ket{\psi_2}$ the optimal success probability $p_d$ is given by the expression
\begin{eqnarray}
p_d=1-|\langle\psi_1|\psi_2\rangle|,
\end{eqnarray}
under the assumption that states $\ket{\psi_1}$ and $\ket{\psi_2}$ are generated with equal a priory probabilities.

Unambiguous state discrimination of two nonorthogonal states can be experimentally realized \cite{Kim,Torres1,Torres2,Neves1} with the help of conditional interferometers $CI_0$ and $CI_1$ illustrated in Fig. \ref{Transformation}. Input and output ports of these interferometers are defined by polarizing beam splitters. Each interferometer has in one arm a half-wave plate which modifies the polarization state of photons conditional on the propagation path.
 
Conditional interferometer $CI_0$ receives photons with spatial state $\ket{0}_1$ at the output of interferometer $I_1$. Interferometer $CI_0$ is described by the transformation $U^{(0)}$ whose action is given by
\begin{eqnarray}
U^{(0)}\ket{h}_{1}\ket{0}_{1} & = & \cos(2\gamma) \ket{h}_{1}\ket{0'}_1 + \sin(2\gamma)\ket{v}_{1}\ket{2'}_1, 
\nonumber\\
U^{(0)}\ket{v}_{1} \ket{0}_{1}& = & e^{i\chi} \ket{v}_{1}\ket{0'}_1,
\label{U0}
\end{eqnarray}
where $\gamma$ is the tilting angle of the half-wave plate inside the interferometer, and states $\ket{0'}_1$ and $\ket{2'}_1$ describe the spatial state of photons emerging from the output ports of the conditional interferometer. The value of $\gamma$ is given by
\begin{eqnarray}
\gamma=\frac{1}{2}\arccos(c_{h}/c_{v}),
\label{Gamma}
\end{eqnarray}
and $\chi=0$. This interferometer implements the transformation of Eq. (\ref{TRANSFORMATION1}). The nonorthogonal states are
given by $|\alpha_{\pm}\rangle_1|0\rangle_1$ of Eq. (\ref{ALFAS}) and these are mapped onto the orthogonal states $(1/\sqrt{2})(|h\rangle_1|0'\rangle_1\pm |v\rangle_1|0'\rangle_1)$. The failure of the discrimination process is indicated by the detection of vertically polarized photons on path $|2'\rangle_1$. The probability $p_{d,0}$ of successfully discriminating between the nonorthogonal states is given by $p_{d,0}=2c_h^2$.

The second conditional interferometer $CI_1$ acts onto photons described by the spatial state $\ket{1}_1$ and implements the transformation of Eq. (\ref{TRANSFORMATION2}).
The action of this interferometer is analogous to the action of $CI_0$, that is it is described by the transformation $U^{(1)}$ given by
 \begin{eqnarray}
U^{(1)}\ket{v}_{1}\ket{1}_{1} & = & \cos(2\gamma) \ket{v}_{1}\ket{1'}_1 + \sin(2\gamma)\ket{h}_{1}\ket{3'}_1, 
\nonumber\\
U^{(1)}\ket{h}_{1} \ket{1}_{1}& = & e^{i\chi} \ket{h}_{1}\ket{1'}_1,
\label{U1}
\end{eqnarray}
where  $\ket{1'}_1$ and $\ket{3'}_1$ describe the spatial state of photons emerging from the output ports of $CI_1$. The nonorthogonal states to be discriminated by $CI_1$
are $|\tilde\alpha_{\pm}\rangle_1|1\rangle_1$ of Eq. (\ref{ALFAST}) and these are mapped onto the orthogonal states $(1/\sqrt{2})(|h\rangle_1|1'\rangle_1\pm |v\rangle_1|1'\rangle_1)$. The failure of the discrimination process is indicated by the detection of vertically polarized photons on path $|3'\rangle_1$. The probability $p_{d,1}$ of successfully discriminating between the nonorthogonal states is given by $p_{d,1}=2c_h^2$. 

The joint action of conditional interferometers $CI_0$ and $CI_1$ onto state $|\tilde\phi^{(2)}\rangle_1$ is thus given by
\begin{eqnarray}
U^{(0)}U^{(1)}|\tilde\phi^{(2)}\rangle_1&=&\frac{c_h}{a\sqrt{2}} 
(a|h\rangle_1\pm b|v\rangle_1)(a|0'\rangle_1\pm b|1'\rangle_1)
\nonumber\\
&+&\frac{\sqrt{c_v^2-c_h^2}}{a\sqrt{2}} (a^2\ket{v}_{1}\ket{2'}_{1}
+ b^2\ket{h}_{1}\ket{3'}_{1} ).
\nonumber\\
\label{Correction}
\end{eqnarray}
The first term at the right hand side of the previous equation describes a perfect cloning process where the clones are encoded as a superposition of the polarization states $\ket{h}_1$ and $\ket{v}_1$ of photon $1$ and as a superposition of the spatial states $\ket{0'}_1$ and $\ket{1'}_1$ of this photon. The second term at the right hand side of Eq. (\ref{Correction}) is a partially entangled state between path and polarization of photon $1$. This state describes the failure of the attempt to correct the errors introduced by the partially entangled quantum channel and does not depend on the state to be cloned.

A photo-detection on path $b$ of photon 2 projects photon 1 to the state $|\tilde\phi^{(3)}\rangle_1$. In this case the errors introduced by the partially entangled quantum channel can also be corrected using transformations $U^{(0)}$ and $U^{(1)}$, as in the case of state $|\tilde\phi^{(2)}\rangle_1$. These two transformations are carried out after applying the recovery unitary operator. The complete sequence of transformations is depicted in Fig. (\ref{Transformation}).

The total probability $p_{total}^{pe}$ of generating faithful clones through a partially entangled quantum chanel is given in this case by the product of the optimal cloning probability with the optimal probability of unambiguous state discrimination, that is
\begin{eqnarray}
p_{total}^{pe}=\frac{1}{2}\left(\frac{1-|\langle\alpha_+|\alpha_-\rangle|}{1+|\langle\psi_+|\psi_-\rangle|}\right),
\end{eqnarray}
where the factor $1/2$ enters in this expression due to the filtering process carried out on photon 1. Thereby, the overall probability of generating faithful clones via a partially entangled state is reduced when compared to probability $p_{total}^{me}$ of a maximally entangled quantum channel.

\section{Nonlocal probabilistic cloning via a partially entangled polarization state}\label{SECTION5}
As we discussed in the last section, it is possible to correct errors induced by a partially entangled state as a resource of a local cloning process. In \cite{Araneda} we propose an experimental setup for implementing optimal non-local probabilistic cloning of two arbitrary nonorthogonal states . The proposed implementation requires a maximally entangled source of photons generated via spontaneous parametric down conversion. In this section we show that the cloning process is still feasible using a partially entangled source of photons, but the process requires an additional step in the experimental setup, corresponding to the same shown in (\ref{Transformation}), and the probability of success is reduced. \\
Following the scheme described in \cite{Araneda}, suppose now instead of a maximally entangled state the process starts with a polarization entangled state
\begin{eqnarray}
\ket{\tilde{\Psi}_{0}}=c_{h}\ket{h}_{1}\ket{h}_{2} + c_{v}\ket{v}_{1}\ket{v}_{2},
\end{eqnarray}
where $c_{h}$ and $c_{v}$ are reals, $c_{h}\leq c_{v}$ and $|c_{h}|^{2}+|c_{v}|^{2}=1$. Following the setup showed in \cite{Araneda} is easy to see that after state codification the state $\ket{\tilde{\Psi}_{0}}$ can be written as
\begin{eqnarray}
\ket{\tilde{\Psi}_{1}} = \ket{\psi_{\pm}}_{1}\left( c_{h}\ket{0}_{1}\ket{h}_{2}\ket{0}_{2}+c_{v}\ket{1}_{1}\ket{v}_{2}\ket{1}_{2}\right).  
\end{eqnarray}
After using several optical elements (PBS's and HWP's) it is possible write the state above as
\begin{eqnarray}
\ket{\tilde{\Psi}_{2}}=\ket{\vartheta} + \ket{\tilde{\vartheta}},
\end{eqnarray}
where
\begin{eqnarray}
\ket{\vartheta} & = & ac_{v}\ket{h}_{1}\ket{1}_{1}(\alpha\ket{h}_{2}\ket{0}_{2}+\beta \ket{v}_{2}\ket{1}_{2}) \nonumber \\
                &   & \pm bc_{h}\frac{1}{\sqrt{2}}\ket{h}_{1}\ket{0}_{1}(\ket{h}_{2}\ket{1}_{2}+\ket{v}_{2}\ket{0}_{2}) \\
\ket{\tilde{\vartheta}} & = & ac_{h}\ket{h}_{1}\ket{\tilde{0}}_{1}\frac{1}{\sqrt{2}}(\ket{h}_{2}\ket{1}_{2}+\ket{v}_{2}\ket{0}_{2}) \nonumber \\
 & & \pm bc_{v}\ket{h}_{1}\ket{\tilde{1}}_{1}(\alpha \ket{h}_{2}\ket{0}_{2}+\beta \ket{v}_{2}\ket{1}_{2}).               
\end{eqnarray}
After the action of two interferometers acting on photon 1 the state of the system can be written as
\begin{eqnarray}
\ket{\tilde{\Psi}_{3}} & = & ac_{v}\tilde{\beta}\ket{v}_{1}\ket{1}_{1}\ket{\tilde{\phi}^{(1)}}_{2}+\frac{i}{\sqrt{2}}\ket{h}_{1}\ket{a}_{1}\ket{\tilde{\phi}^{(2)}}_{2} \nonumber \\
& &  +\frac{1}{\sqrt{2}}\ket{h}_{1}\ket{b}_{1}\ket{\tilde{\phi}^{(3)}}_{2}+ac_{h}\tilde{\beta}\ket{v}_{1}\ket{\tilde{0}}_{1}\ket{\tilde{\phi}^{(4)}}_{2} \nonumber \\ 
& & +\frac{i}{\sqrt{2}}\ket{h}_{1}\ket{c}_{1}\ket{\tilde{\phi}^{(5)}}_{2}+\frac{1}{\sqrt{2}}\ket{h}_{1}\ket{d}_{1}\ket{\tilde{\phi}^{(6)}}_{2},
\end{eqnarray}
in which each state $\ket{\tilde{\phi}^{(i)}}_{2}$ represents a different detection path for photon 1, 
\begin{eqnarray}
\ket{\tilde{\phi}^{(1)}}_{2} & = & \alpha\ket{h}_{2}\ket{0}_{2}+\beta\ket{v}_{2}\ket{1}_{2},\\
\ket{\tilde{\phi}^{(2)}}_{2} & = & ac_{v}\tilde{\alpha}\left(\alpha\ket{h}_{2}\ket{0}_{2}+\beta\ket{v}_{2}\ket{1}_{2}\right)\nonumber \\
& & \pm bc_{h}\frac{1}{\sqrt{2}}\left(\ket{h}_{2}\ket{1}_{2}+\ket{v}_{2}\ket{0}_{2}\right),\\
\ket{\tilde{\phi}^{(3)}}_{2} & = & ac_{v}\tilde{\alpha}\left(\alpha\ket{h}_{2}\ket{0}_{2}+\beta\ket{v}_{2}\ket{1}_{2}\right)\nonumber \\
& & \mp bc_{h}\frac{1}{\sqrt{2}}\left(\ket{h}_{2}\ket{1}_{2}+\ket{v}_{2}\ket{0}_{2}\right),\\
\ket{\tilde{\phi}^{(4)}}_{2} & = & \frac{1}{\sqrt{2}}\left(\ket{h}_{2}\ket{1}_{2}+\ket{v}_{2}\ket{0}_{2}\right),\\
\ket{\tilde{\phi}^{(5)}}_{2} & = & ac_{h}\tilde{\alpha}\frac{1}{\sqrt{2}}\left(\ket{h}_{2}\ket{1}_{2}+\ket{v}_{2}\ket{0}_{2}\right)\nonumber \\
& & \pm bc_{v}\left(\alpha\ket{h}_{2}\ket{0}_{2}+\beta\ket{v}_{2}\ket{1}_{2}\right),\\
\ket{\tilde{\phi}^{(6)}}_{2} & = & ac_{h}\tilde{\alpha}\frac{1}{\sqrt{2}}\left(\ket{h}_{2}\ket{1}_{2}+\ket{v}_{2}\ket{0}_{2}\right)\nonumber \\
& & \mp bc_{v}\left(\alpha\ket{h}_{2}\ket{0}_{2}+\beta\ket{v}_{2}\ket{1}_{2}\right).
\end{eqnarray}
Then, according to the outcome measurement of photon 1 the cloning process could be successful or not. 
In case photon 1 is detected in path $\ket{1}_{1}$ or $\ket{\tilde{0}}_{1}$ the state of photon 2 is projected onto state $\ket{\tilde{\phi}^{(1)}}$ and $\ket{\tilde{\phi}^{(4)}}$ respectively, therefore, the process fails. On the other hand, if photon 1 is detected, for example, in path $\ket{a}_{1}$ the state of photon 2 will be projected to the state $\ket{\tilde{\phi}^{(2)}}$, which can be written as
\begin{eqnarray}
\ket{\tilde{\phi}^{(2)}}_{2}&=&\frac{1}{\sqrt{2}}\left[a\left(c_{v}a\ket{h}_{2}\pm c_{h}b\ket{v}_{2}\right)\ket{0}_{2}\right.  \nonumber \\
& & \left. \pm b\left(c_{h}a\ket{h}_{2}\pm c_{v}b\ket{v}_{2}\right)\ket{1}_{2}\right]. \label{phi2}
\end{eqnarray}
According to this equation it is possible to see that coefficients $c_{h}$ and $c_{v}$ lead an error in the cloned states, nevertheless this error can be corrected using techniques of quantum state discrimination. To do this lets define the two pairs of non-orthogonal states
\begin{eqnarray}
\ket{\alpha_{\pm}}_{2}\ket{0}_{2} & = & c_{v}\ket{h}_{2}\ket{0}_{2}\pm c_{h}\ket{v}_{2}\ket{0}_{2,}\label{eq:alpha-}\\
\ket{\tilde{\alpha}_{\pm}}_{2}\ket{1}_{2} & = & c_{h}\ket{h}_{2}\ket{1}_{2}\pm c_{v}\ket{v}_{2}\ket{1}_{2}.\label{eq:alpha-tilde}
\end{eqnarray}
Now it is possible to write state $\ket{\tilde{\phi}^{(2)}}$ as
\begin{eqnarray}
\ket{\tilde{\phi}^{(2)}}_{2}&=&\frac{1}{\sqrt{2}a}\left\{ a\left[d_{\pm}\ket{\alpha_{+}}_{2}+d_{\mp}\ket{\alpha_{-}}_{2}\right]\ket{0}_{2}\right.  \nonumber \\
& & \left.  \pm b\left[d_{\pm}\ket{\tilde{\alpha}_{+}}_{2}+d_{\mp}\ket{\tilde{\alpha}_{-}}_{2}\right]\ket{1}_{2}\right\} ,
\end{eqnarray}
where $d_{\pm}=(a\pm b)/2$. As was shown in section (\ref{SECTIONIV}), making use of Unambiguous State Discrimination is possible to perform the following transformations, 
\begin{eqnarray}
\ket{\alpha_{\pm}}_{2}\ket{0}_{2} & \rightarrow & \frac{1}{\sqrt{2}}\left(\ket{h}_{2}\ket{0}_{2}\pm\ket{v}_{2}\ket{0}_{2}\right),\label{eq:transfor-1}\\
\ket{\tilde{\alpha}_{\pm}}_{2}\ket{1}_{2} & \rightarrow & \frac{1}{\sqrt{2}}\left(\ket{h}_{2}\ket{1}_{2}\pm\ket{v}_{2}\ket{1}_{2}\right),\label{eq:transfor-2}
\end{eqnarray}
which correct the state $\ket{\tilde{\phi}^{(2)}}$, however this transformation also introduces a non-conclusive result in the process which is characterized by a certain probability of success. 
%In general, in the case of two states $\ket{\psi_{1}}$ and $\ket{\psi_{2}}$ are generated with the same prior probability, the probability of unambiguously discriminate between them is given by
%\begin{eqnarray}
%p_{d}=1-\left|\langle\psi_{1}|\psi_{2}\rangle\right|.
%\end{eqnarray}
This transformation can be implemented using a Conditional Interferometer. In order to correct the state (\ref{phi2}) two interferometers are needed, one for the pair (\ref{eq:alpha-}) and other for (\ref{eq:alpha-tilde}), see (\ref{Transformation})
The first interferometer, $CI_{0}$, acts on photons coming from path $\ket{0}_{2}$ and is characterized by the transformation
\begin{eqnarray}
U^{(0)}\ket{h}_{2}\ket{0}_{2}=\cos\left(2\gamma\right)\ket{h}_{2}\ket{0'}_{2}+\sin\left(2\gamma\right)\ket{v}_{2}\ket{2'}_{2},
\end{eqnarray}
where $\gamma$ is the angle defined by the half wave plate inside the interferometer and has to be set to
\begin{eqnarray}
\gamma=\frac{1}{2}\arccos\frac{c_{h}}{c_{v}}.
\end{eqnarray}
The paths $\ket{0'}_{2}$ and $\ket{2'}_{2}$ define the result of the transformation, that means whether discrimination was successful, $\ket{0'}_{2}$, or failed, $\ket{2'}_{2}$. Finally, the probability of successfully discriminate the pair of states (\ref{eq:alpha-}) is given by
\begin{eqnarray}
p_{d,0}=2c_{h}^{2}.
\end{eqnarray}
The procedure is similar for the other interferometer, $CI_{1}$, where the transformation is characterized by 
\begin{eqnarray}
U^{(1)}\ket{v}_{2}\ket{1}_{2}=\cos\left(2\gamma\right)\ket{v}_{2}\ket{1'}_{2}+\sin\left(2\gamma\right)\ket{h}_{2}\ket{3'}_{2},
\end{eqnarray}
and the probability of success is given by
\begin{eqnarray}
p_{d,1}=2c_{h}^{2}.
\end{eqnarray}
The joint action of $CI_{0}$ and $CI_{1}$ over state $\ket{\tilde{\phi}^{(2)}}_{2}$ is given by
\begin{eqnarray}
U^{(0)}U^{(1)}\ket{\tilde{\phi}^{(2)}}_{2}&=&\frac{c_{h}}{a\sqrt{2}}\left(a\ket{h}_{2}\pm b\ket{v}_{2}\right)\left(a\ket{0'}_{2}\pm b\ket{1'}_{2}\right)\nonumber \\
& & \pm\frac{\sqrt{c_{v}^{2}-c_{h}^{2}}}{a\sqrt{2}}\left(a^{2}\ket{v}_{2}\ket{2'}_{2}+b^{2}\ket{h}_{2}\ket{3'}_{2}\right)\nonumber.\\\label{eq:result}
\end{eqnarray}
The first term describes the successful process, where the two clones are generated in the polarization and spatial degree of freedom of photon 2, and the last term of equation (\ref{eq:result}) describes the failure process in which an entangled state is obtained, this state is independent of the state to be cloned. 
The process is the same for state $\ket{\tilde{\phi}^{(3)}}_{2}$, and is analogous for the remaining states, $\ket{\tilde{\phi}^{(5)}}_{2}$ and $\ket{\tilde{\phi}^{(6)}}_{2}$, in which the tilting angle of the half wave plate in the interferometers has to be changed and an extra unitary transformation (RUO) is necessary before applying the two conditional interferometers.
Finally, the total probability of success is determined by both the optimum probability of cloning and the probability of unambiguously discriminate between two non-orthogonal states,  
\begin{eqnarray}
p_{total}^{pe}=\frac{1-\left|\langle\alpha_{+}|\alpha_{-}\rangle\right|}{1+\left|\langle\psi_{+}|\psi_{-}\rangle\right|}=\frac{2c_{h}^{2}}{1+\cos\theta}.
\end{eqnarray}
\section{Summary} 
\label{SECTIONV}

The cloning of quantum states has been a constant subject of study over the last decades. This process has been experimentally implemented in the case of approximate universal cloning \cite{Cummins,Xu,Lamas,Martini,Fasel,Ricci,Irvine,Huang,Howell,Fiurasek,Sciarrino,Pan} and variations. The realization of probabilistic cloning has been only recently achieved via nuclear magnetic resonance \cite{Hongwei}. Since this proposal requires the use of ensemble techniques, it does not constitute cloning of individual quantum systems.

Here, we have studied the possibility of implementing probabilistic cloning via twin photons in an all-optical setup. This allows the broadcasting of the clones. We have proposed an experimental setup for realizing the probabilistic cloning of two nonorthogonal quantum states. The states to be cloned are encoded in the polarization degree of freedom of a single photon while the clones are encoded in the two available propagation paths of the same photon. The resting twin photon is employed as an ancillary system to carry out conditional transformations. This proposal achieves an overall success probability equal to the half of the optimal cloning probability. This is due to the use of a filtering process after that the state to be cloned has been encoded. A second characteristic of our proposal is the use of recovery operators to correct errors arisen in the cloning process itself, which is akin to quantum teleportation \cite{Lombardi,Boschi}.
A different approach to the problem of probabilistic nonlocal cloning  has also been reported \cite{Zhang} but considering a GHZ state of a tripartite system as initial source and make use of an additional port qubit owned by the sender which make it more difficult to implement. Additional, it is not symmetric, which means that each copy is created with a different success  probability. 

We have also studied the consequences of a partially entangled state as quantum channel for the cloning process and shown that the errors generated by this class of channel can be probabilistically corrected by means of unambiguous state discrimination. The total probability of generating perfect clones is then reduced by a factor equal to the optimal probability of discriminating two nonorthogonal states which are defined by the Schmidt coefficients of the quantum channel. In this way we showed that the effect of having a nonmaximal entangled state as a source and correcting after is a diminution in the rate of the perfect clones generation.

\section{Acknowledgments}

This work was supported by Grant MSI P010-30F and Grants FONDECYT No. 11085057 and No. 11121318.

%\section{}
%\label{}

%% The Appendices part is started with the command \appendix;
%% appendix sections are then done as normal sections
%% \appendix

%% \section{}
%% \label{}

%% References
%%
%% Following citation commands can be used in the body text:
%% Usage of \cite is as follows:
%%   \cite{key}          ==>>  [#]
%%   \cite[chap. 2]{key} ==>>  [#, chap. 2]
%%   \citet{key}         ==>>  Author [#]

%% References with bibTeX database:

%\bibliographystyle{model1-num-names}
%\bibliography{<your-bib-database>}

\section*{References}
\begin{thebibliography}{00}

%% \bibitem must have the following form:
%%   \bibitem{key}...
%%

% \bibitem{}
%Teleportation

\bibitem{Wootters}W. K. Wootters and W. H. Zurek, Nature (London) {\bf 299}, 802 (1982).
\bibitem{Dieks} D. Dieks, Phys. Lett. A {\bf 92}, 271 (1982).
\bibitem{Milonni} P. W. Milonni and M. L. Hardies, Phys. Lett. A {\bf 92}, 321 (1982).
\bibitem{Mandel} L. Mandel, Nature (London) {\bf 304}, 188 (1983).
\bibitem{Buzek} V. Bu\v zek and M. Hillery, Phys. Rev. A {\bf 54}, 1844 (1996).

\bibitem{Rev} Heng Fan, Yi-Nan Wang, Li Jing,  Jie-Dong Yue, Han-Duo Shi, Yong-Liang Zhang and Liang-Zhu Mu, Phys. Rep. {\bf 44} 241 (2014)

\bibitem{photons} A. Fedrizzi, T. Herbst, A. Poppe, T. Jennewein and A. Zeilinger, Optics Express {\bf 15}, 15377 (2007)


\bibitem{Werner} R. F. Werner, Phys. Rev. A \textbf{58}, 1827 (1998).
\bibitem{Keyl} M. Keyl and R. F. Werner, J. Math. Phys. \textbf{40}, 3283 (1999).
\bibitem{Alber} G. Alber, A. Delgado and I. Jex, Quantum Inf. and Comput. {\bf 1}, 33 (2001). 
\bibitem{Scarani} V. Scarani, S. Iblisdir, N. Gisin, and A. Ac\'in, Rev. Mod. Phys. \textbf{77}, 1225 (2005).
\bibitem{Duan} Lu-Ming Duan and Guang-Can Guo, Phys. Lett. A \textbf{243}, 261 (1998).
\bibitem{Duan2} Lu-Ming Duan and Guang-Can Guo, Phys. Rev. Lett. \textbf{80}, 4999 (1998).
\bibitem{Hongwei} Hongwei Chen, Dawei Lu, Bo Chong, Gan Qin, Xianyi Zhou, Xinhua Peng, and Jiangfeng Du, Phys. Rev. Lett. {\bf 106}, 180404 (2011).
\bibitem{Araneda} G. Araneda, N. Cisternas, O. Jim\'enez, and A. Delgado, Phys. Rev. A {\bf 86}, 052332 (2012).
\bibitem{DelgadoC} L. Neves, M. A. Sol\'is-Prosser, A. Delgado, and O. Jim\'enez, Phys. Rev. A {\bf 85}, 062322 (2012). 
\bibitem{DelgadoD} L. Roa, A. Delgado, and I. Fuentes-Guridi, Phys. Rev. A {\bf 68}, 022310 (2003).
\bibitem{DelgadoA} O. Jim\'enez, J. Bergou, and A. Delgado, Phys. Rev. A {\bf 82}, 062307 (2010).
\bibitem{DelgadoB} O. Jim\'enez, L. Roa, and A. Delgado, Phys. Rev. A {\bf 82}, 022328 (2010).

\bibitem{Ivanovic87} I. D. Ivanovic, Phys. Lett. A {\bf 123}, 257 (1987); D. Dieks, {\it ibid.} {\bf 126}, 303 (1988); A. Peres, {\it ibid.} {\bf 128}, 19 (1988); G. Jaeger and A. Shimony, {\it ibid.} {\bf 197}, 83 (1995).
\bibitem{Chefles98-2} A. Chefles, Phys. Lett. A {\bf 239}, 339 (1998).
\bibitem{Chefles98-1} A. Chefles and S. M. Barnett, Phys. Lett. A {\bf 250}, 223 (1998).

\bibitem{Croke06} S. Croke, E. Andersson, S. M. Barnett, C. R. Gilson, and J. Jeffers, Phys. Rev. Lett.\ \textbf{96}, 070401 (2006).

\bibitem{DelgadoE} O. Jim\'enez, X. S\'anchez-Lozano, E. Burgos-Inostroza, A.
Delgado, and C. Saavedra, Phys. Rev. A {\bf 76}, 062107 (2007); O.	Jim\'enez, M. A. Sol\'is-Prosser, A. Delgado, and L. Neves, {\it ibid.} {\bf 84}, 062315 (2011); U. Herzog, {\it ibid.} {\bf 85}, 032312 (2012).

\bibitem{Kim} Yoon-Ho Kim, Phys. Rev. A {\bf 67}, 040301R (2003).
\bibitem{Torres1} F. A. Torres-Ruiz, J. Aguirre, A. Delgado, G. Lima, L. Neves, S. P\'adua, L. Roa, and C. Saavedra, Phys. Rev. A {\bf 79}, 052113, (2009).

\bibitem{Torres2} F. A. Torres-Ruiz, G. Lima, A. Delgado, S. P\'adua, and C. Saavedra, Phys. Rev. A {\bf 81}, 042104 (2010).
\bibitem{Neves1} L. Neves, G. Lima, J. Aguirre, F. A. Torres-Ruiz, C. Saavedra, and A. Delgado, New J. Phys. {\bf 11}, 073035 (2009).

\bibitem{B92} C.H. Bennett, Phys. Rev. Lett., {\bf 68} 3121 (1992).

\bibitem{appli} E. Pomarico, B. Sanguinetti, P. Sekatski, H. Zbinden,
and N. Gisin, Optics and Spectroscopy {\bf 111}, 510 (2011)


\bibitem{Cummins} H. K. Cummins, C. Jones, A. Furze, N. F. Soffe, M. Mosca, J. M. Peach, and J. A. Jones, Phys. Rev. Lett. {\bf 88}, 187901 (2002).
\bibitem{Xu} Jin-Shi Xu, Chuan-Feng Li, Lei Chen, Xu-Bo Zou, and Guang-Can Guo, Phys. Rev. A {\bf 78}, 032322 (2008).
\bibitem{Lamas} A. Lamas-Linares, C. Simon, J. Howell, and D. Bouwmeester, Science {\bf 296}, 712 (2002).
\bibitem{Martini} F. De Martini, V. Bu\v zek, F. Sciarrino, and C. Sias, Nature (London) {\bf 419}, 815 (2002).
\bibitem{Fasel} S. Fasel, N. Gisin, G. Ribordy, V. Scarani, and H. Zbinden, Phys. Rev. Lett. {\bf 89}, 107901 (2002).
\bibitem{Ricci} M. Ricci, F. Sciarrino, C. Sias, and F. De Martini, Phys. Rev. Lett. {\bf 92}, 047901 (2004).
\bibitem{Irvine} W. T. M. Irvine, A. L. Linares, M. J. A. de Dood, and D. Bouwmeester, Phys. Rev. Lett. {\bf 92}, 047902 (2004).
\bibitem{Huang} Y.-F. Huang, W.-L. Li, C.-F. Li, Y.-S. Zhang, Y.-K. Jiang, and G.-C. Guo, Phys. Rev. A {\bf 64}, 012315 (2001).
\bibitem{Howell} I. Ali Khan and J. C. Howell, Phys. Rev. A {\bf 70}, 010303R (2004).
\bibitem{Fiurasek} A. \v Cernoch, L. Bartu\v skov\'a, J. Soubusta, M. Je\v zek, J. Fiur\'a\v sek, and M. Du\v sek, Phys. Rev. A {\bf 74}, 042327 (2006).
\bibitem {Sciarrino} F. Sciarrino and F. De Martini, Phys. Rev. A {\bf 72}, 062313 (2005).
\bibitem{Pan} Xin-Yu Pan, Gang-Qin Liu, Li-Li Yang, and Heng Fan, Appl. Phys. Lett. {\bf 99}, 051113 (2011).

\bibitem{Zhang} Chuan-Wei Zhang, Chuan-Feng Li, Zi-Yang Wang, and Guang-Can Guo, Phys. Rev. A {\bf 62}, 042302 (2000).
\bibitem{Lombardi} E. Lombardi, F. Sciarrino, S. Popescu, and F. De Martini, Phys. Rev. Lett. {\bf 88}, 070402 (2002).
\bibitem{Boschi} D. Boschi, S. Branca, F. De Martini, L. Hardy, and S. Popescu, Phys. Rev. Lett. {\bf 80}, 1121 (1998).

\bibitem{Kwiat} P. G. Kwiat, E. Waks, A. G. White, I. Appelbaum and P. H. Eberhard, Phys. Rev. A {\bf 60}, R773(R) (1999)

\end{thebibliography}

%% Authors are advised to submit their bibtex database files. They are
%% requested to list a bibtex style file in the manuscript if they do
%% not want to use model1-num-names.bst.

%% References without bibTeX database:

\end{document}